\documentstyle[aps,prb,epsfig]{revtex}

\begin{document}
\draft
\author{Tomosuke Aono$^1$ and
Mikio Eto$^{2,3}$}
\address{$^1$The Institute of Physical and Chemical Research (RIKEN),\\
2-1 Hirosawa, Wako-shi, Saitama 351-0198 Japan\\
$^2$
Faculty of Science and Technology, Keio
University,\\ 3-14-1 Hiyoshi,
Kohoku-ku, Yokohama 223-8522 Japan\\
$^3$Department of Applied Physics/DIMES,
Delft University of Technology,\\
Lorentzweg 1, 2628 CJ Delft,
The Netherlands}
\title{Kondo resonant spectra in coupled quantum dots}
\date{\today}
\maketitle
\begin{abstract}
The Kondo effect in coupled quantum dots
is investigated
from the viewpoint of transmission spectroscopy
using the slave-boson formalism of the Anderson model.
The antiferromagnetic spin-spin coupling $J$ between the dots is
taken into account.
Conductance $G$ through the dots connected in a series is characterized by
the competition between the dot-dot tunneling coupling $V_{C}$ and the
level broadening $\Delta$ in
the  dots (dot-lead coupling).
When $V_{C}/\Delta < 1$, the Kondo resonance is formed between each dot and lead,
which is replaced by a spin-singlet state in the dots
at low gate voltages.
The gate voltage dependence of $G$ has a sharp peak
of  $2 e^2/h$ in height in the crossover region between the Kondo and
spin-singlet states.
The sharp peak of $G$
survives when the energy levels are different between the dots.
When $V_{C} / \Delta > 1$, the ``molecular levels" between the Kondo resonant states appear;
the Kondo resonant peaks are located
below and above the Fermi level in the leads at low gate voltages.
The gate voltage dependence of $G$ has a broad peak,
which is robust against $J$.
The broad peak splits into two peaks when the energy levels are different,
reflecting the formation of the asymmetric molecular levels between the Kondo resonant states.
\end{abstract}

\section{Introduction}
\label{sec1}

Recently the Kondo effect has been observed in semiconductor quantum dots
connected to external leads by tunneling barriers.
\cite{Goldhaber,Cronenwett,Schmid98,Simmel99} The Kondo effect
makes a resonant state at the Fermi level in the leads when the number of electrons
in the dot is odd.
\cite{Glazman,Ng,Kawabata,Hershfield,Meir,Yeyati93,Oguri95,Konig}
The resonant width is given by the Kondo temperature $T_K$.
This results in (i) the unitary limit of the linear conductance through
the dot $G=2e^2/h$ at low temperatures ($T \ll T_K$) (Ref.~\onlinecite{Glazman}) and
(ii) the zero-bias peak of the differential conductance $dI/dV_{\rm sd}$ with the
width of $T_K$ under finite source-drain voltages.~\cite{Meir}

In this paper, we theoretically examine the Kondo resonant state in double
quantum dots connected in a series, as will be shown in Fig.\ 1.
We consider one level in each quantum dot, often referred to as
an ``artificial atom".~\cite{Tarucha,METbook}
At $T \gg T_K$, the transport properties of such systems
have been investigated by several experiments,~\cite{Austing,Waugh,Blick,Oosterkamp}
from a view point of the formation of an ``artificial molecule." The molecular
orbitals between the two dots have been observed,~\cite{Blick,Oosterkamp}
which reflect the coherent coupling between the dots ($V_{C}$ in Fig.\ 1).
When the double dot system accommodates two electrons,
the coherent coupling and intra-dot interaction $U$ make many-body
states \cite{HL,eto}. The ground state is a spin singlet. The excitation
energy to a spin-triplet state is given by
\begin{equation}
J=\frac{V_{C}}{2} \left[ \sqrt{(U/V_{C})^2+16}-(U/V_{C})
\right] \approx 4 V_{C}^2/U,
\label{eq:J_def}
\end{equation}
when $U \gg V_{C}$. The ground state is an ``entangled" state
between two localized spins,
which has attracted new interest for the application to quantum
computing.~\cite{Loss}
At $T \ll T_{K}$,
the two elements $V_{C}$ and $J$ would lead to rich structures of the
Kondo
resonant state in this system.

A useful tool to elucidate the Kondo resonance is
the slave-boson formalism of the Anderson
model with $U\rightarrow
\infty$.~\cite{Read83,Coleman87,Bickers87,Newns88,HewsonBook} In the case of a
single quantum dot, the dot state is spin-up
$f_{\uparrow}^{\dagger} |0\rangle$, spin-down $f_{\downarrow}^{\dagger}
|0\rangle$,
or empty $b^{\dagger} |0\rangle$
($f_{\uparrow}^{\dagger}f_{\uparrow}+f_{\downarrow}^{\dagger}f_{\downarrow}+
b^{\dagger}b=1$).
In the mean-field approximation,
$b$ and $b^\dagger$ are replaced by
a real $c$-number, $b^{\dagger} = b = b_0$.
The density of states for pseudo fermions per spin is
\cite{electron_dos}
\begin{equation}
\rho_{f}(\omega) =   - \frac{{\rm Im} }{\pi} \frac{1}{\omega - \widetilde{E} + {\rm i}
\widetilde{\Delta}}
             =  \frac{1}{\pi} \frac{\widetilde{\Delta}}{ ( \omega - \widetilde{E})^2 +
\widetilde{\Delta}^2}.
\end{equation}
In the Kondo regime where one electron exists in the dot
($b_0^2\ll 1$, $f_{\uparrow}^{\dagger}f_{\uparrow}+f_{\downarrow}^{\dagger}f_{\downarrow} \cong 1$),
the density of states $\rho_f(\omega)$ is half filled and, hence, the center of the
resonant state
$\widetilde{E}$ is matched with the Fermi level. The width $\widetilde{\Delta}$ is the Kondo
temperature
\begin{equation}
T_{{K},1} = \frac{\Delta}{\pi} \exp ( \pi E^*/\Delta),
\label{eq:Kondo_temperature}
\end{equation}
where $\Delta$ is the level broadening, $\pi \rho |V|^2$,
with the density of states $\rho$ in the external leads and the dot lead
tunneling coupling $V$.
$E^*$ is the renormalized dot-level,
$E^{*} =  V_{g} + \Delta/\pi
\ln \left( D \pi/\Delta \right) $
with the gate voltage $V_g$ and the bandwidth $D$ in the leads.~\cite{Haldane78}
Therefore this method is suitable to examine how the Kondo resonant
state is formed.
When the number of electrons in the dot is less than unity
(valence fluctuating regime),
the center of the resonance moves above the Fermi level.
The resonant width increases from $T_K$
to $\Delta$,
as the number of electrons in the dot decreases.
However,
the mean-field approximation is
not quantitatively accurate in the valence fluctuating regime.

The Kondo effect on the transport properties in coupled
dot systems have been studied by several theoretical
methods.~\cite{Izumida,Ivanov,Pohjola,Andrei,Busser99,Izumida99}
In our previous
paper,~\cite{Aono98} we have applied the slave boson mean-field theory to the
double dot
system
in the absence of the antiferromagnetic spin-spin coupling $J$.
We have shown that the Kondo resonance is determined by the
competition between the dot-dot tunneling coupling $V_{C}$ and dot-lead
coupling $\Delta$.
The resonant state has two peaks below
and above the Fermi level for $V_{C}/\Delta>1$, whereas it has a single peak
at the Fermi level
for $V_{C}/\Delta<1$. In consequence, the linear-conductance $G$ through the
double dots as a function of the gate voltage
is qualitatively different between the two cases.

Extending our calculations, Georges and Meir\cite{Georges}
have pointed out the
importance
of the spin-spin coupling $J$.
At low-gate voltages, the coupling becomes
relevant
and makes a spin-singlet state in the two dots. Then the Kondo resonance
between
dots and leads is destroyed.
Izumida and Sakai\cite{Izumida99} have examined the same problem
using the numerical renormalization group (NRG) method.
A similar situation was investigated for dilute magnetic systems,
which is known as the two impurity Kondo
problem.~\cite{Jones,Jones1,Jones2,Sakai90,Sakai92,Affleck92}
When the antiferromagnetic coupling $J$,
 the origin of which is the Ruderman-Kittel-Kasuya-Yosida interaction,
increases from zero,
electronic states of two magnetic impurities
exhibit the first-order phase transition
from the Kondo state to the spin-singlet state
at $J \sim T_{K,1}$.
The magnetic susceptibility diverges
at the transition.
The phase transition changes into a crossover in
the case of $V_{C} \neq 0$.~\cite{Sakai90,Sakai92}
In coupled quantum dots,
$J$ is given by Eq. (\ref{eq:J_def})
whereas $T_{K,1}$ increases as the gate voltage increases
as in Eq.\ (\ref{eq:Kondo_temperature}).

As indicated by the previous work \cite{Izumida99,Georges}, the
interplay of $\Delta$, $V_{C}$, and $J$ results in various transport
properties. They are observable by experiments since these parameters
can be controlled in quantum dot systems. The physical origins of the
transport properties, however, are difficult to understand by the NRG
method \cite{Izumida99} and by the conventional argument using the phase
shift.~\cite{Georges}
In this paper,
we look directly at the Kondo resonant state
in the presence of spin-spin coupling $J$.
The effect of $J$ can be understood in terms of
the transmission spectrum,
just as in our previous work in the absence of $J$.
We will show that even for $J \neq 0$,
electron transport is characterized by the competition
between the dot-dot tunneling coupling $V_{C}$ and
the dot-lead coupling $\Delta$.
When $V_C / \Delta < 1$,
the coupling $J$ is essentially important.
$G$ has a sharp peak of $2e^2/h$ in height,
as a function  of gate voltage,
which stems from the coexistence of
the Kondo coupling and spin-singlet coupling.
When $V_C / \Delta > 1$,
the transport properties are robust against $J$.
These results are explained
by only one criterion,
whether the transmission spectrum has double peaks.
Hence our argument is much simpler than the one
based on the phase shift.~\cite{Georges}
In addition,
the transmission spectrum by itself is directly
observable in the differential conductance under finite source-drain voltages,
as suggested by
Aguado and Langreth.~\cite{Aguado99}
Further, we generalize our calculations
to
asymmetric double dot systems.
Some preliminary results have been reported in Ref.~\onlinecite{Aono00}.

The organization of this paper is as follows.
In Sec. \ref{sec2},
the model Hamiltonian for a coupled quantum dot system is introduced.
We explain
the slave boson mean-field theory and derive
the expression of the conductance through the dots.
In Sec. \ref{sec3},
we briefly review
electron transport through symmetric double dots
in the absence of $J$.
The transmission spectra are discussed in detail.
In Sec. \ref{sec4},
we investigate electron transport
through coupled dots in the presence of $J$.
In Sec. \ref{sec5},
the electron transport through asymmetric double dots is examined.
Conclusions and discussion follow in Sec. \ref{sec6}.

\section{Model and Method}
\label{sec2}

\subsection{Model}
Let us consider coupled quantum dots as shown in Fig. \ref{fig1}.
Two dots couple to each other with $V_{C}$,
and to external leads with $V$.
Each dot has a single energy level $E_{\alpha} \;
(\alpha={\rm L,R})$.
We denote the energy difference, $E_L-E_R$, by $\Delta E$.
A common gate $V_{g}$ is attached to
the dots to control $E_{\alpha}$;
we define $V_g = (E_L+E_R)/2$.
We assume that the intradot Coulomb interaction $U$ is
sufficiently large so that
(i) the double occupancy of electrons in each dot is forbidden, but
(ii) the antiferromagnetic spin coupling exists
between the quantum dots due to the virtual double occupancy
in a dot, $J {\bbox S}_{L} \cdot  {\bbox S}_{R}$ with $J$ ($=4 V_{C}^2/U$)
[see Eq. (\ref{eq:J_def})] and
the spin operator ${\bbox S}_{\alpha} (\alpha={L,R})$.
The interdot Coulomb interaction is neglected.
The temperature is set to be zero.

We adopt
the $(N=2)$-fold degenerate Anderson model with
the antiferromagnetic SU($N$) spin interactions,
represented by the slave-boson formalism.~\cite{Georges,Jones2}
We introduce the slave boson operator $b^{\dag}_{\alpha}$,
which creates an empty state
and fermion operator $f_{\alpha m}^{\dagger}$,
which creates a singly occupied state
with spin $m=-j,-(j-1),...,j-1,j (2j+1 = N)$
 in dot $\alpha = L, R$,
under the constraint of
$\widehat{Q}_{\alpha}\equiv
n_{\alpha} +  b^{\dag}_{\alpha} b_{\alpha}=1$ with
$n_{\alpha} =
\sum_{m} f^{\dag}_{\alpha m} f_{\alpha m}$.
The annihilation operator of an electron in dot $\alpha$
is rewritten as
$C_{\alpha m} =  b^{\dag}_{\alpha} f_{\alpha
m}$.~\cite{Read83,Coleman87,Bickers87,Newns88,HewsonBook}

The Hamiltonian is
\begin{equation}
 {\cal H} =
 {\cal H}_0 + {\cal H}_{\rm dot-lead} + {\cal H}_{\rm dot-dot} + {\cal H_{\it
J}}+
\sum_{\alpha = {\rm L, R}} \lambda_{\alpha} ( \hat{Q}_{\alpha} - 1 ).
\label{eq:DefH}
\end{equation}
The first term on
the right-hand side of Eq.\ (\ref{eq:DefH}) represents
electrons in the leads and dots;
\begin{equation}
{\cal H}_0 =
\sum_{\stackrel{\alpha={\rm L, R}}{k, \; m}}
 E(k) c^{\dag}_{\alpha k m}c_{\alpha k m}
+
\sum_{\stackrel{\alpha={\rm L, R}}{m} }
 E_{\alpha} f^{\dag}_{\alpha m} f_{\alpha m}.
\label{eq:HFree}
\end{equation}
The operator $c^{\dag}_{\alpha k m}$ creates an electron in
lead $\alpha$  
with
energy $E(k)$ and spin $m$.
The second and third terms on the right hand side of Eq.\ (\ref{eq:DefH})
describe the dot-lead tunneling coupling and
the dot-dot tunneling coupling, respectively
\begin{equation}
{\cal H}_{\rm dot-lead} =
\frac{V}{\sqrt{N}}  
\sum_{\alpha, k ,m }
c^{\dag}_{\alpha k m}
f_{\alpha m} b^{\dag}_{\alpha} + {\rm H.c.},
\label{eq:HTunnel}
\end{equation}
\begin{equation}
{\cal H}_{\rm dot-dot} =
\frac{V_{C}}{N}
\sum_{m} \;
b^{\dag}_{L} f_{{L} m} f^{\dag}_{{R} m} b_{R} +
{\rm H.c.}
\label{eq:HMix}
\end{equation}
The term $H_{J}$ gives the antiferromagnetic spin-spin interaction
\begin{equation}
{\cal H}_{\it J} =
\frac{J}{N}
\sum_{m,n} \;
f^{\dag}_{{R} m}  f_{{L} m}
f^{\dag}_{{L} n}  f_{{R} n}
+ {\rm H.c.}
\label{eq:HJ}
\end{equation}
The constraint of $\widehat{Q}_{\alpha}=1$ is taken into account
by the last term on the right-hand side of Eq.\ (\ref{eq:DefH})
with the Lagrange multiplier $\lambda_{\alpha}$.

\subsection{Mean-Field Approximation}
Let us now make the following two assumption to treat the Hamiltonian (\ref{eq:DefH})
within the mean-field theory.~\cite{Coleman87,Jones2}
First, the slave-boson operators $b_{\alpha}$ and $b^{\dagger}_{\alpha}$ are replaced by a constant
c-number
$b_{\alpha}$.
This approximation is exact in the limit of $N \rightarrow \infty$ when $J=0$;
it corresponds to
the calculation to the lowest order in the 1/{\it N}
expansion.~\cite{Read83,Coleman87,Bickers87}
Second, we also treat the spin-spin coupling (\ref{eq:HJ})
by the mean-field theory.
It is decoupled by
a parameter $\kappa$,
\begin{equation}
\kappa = \frac{J}{N} \sum_m \langle
f^{\dag}_{{R} m} f_{{L} m} \rangle,
\end{equation}
which characterizes the antiferromagnetic order between the spins in
the dots.
Then we obtain
the following mean-field Hamiltonian;
\begin{eqnarray}
{\cal H}_0 &=&
\sum_{\stackrel{\alpha={\rm L, R}}{k, \; m}}
 E(k) c^{\dag}_{\alpha k m}c_{\alpha k m}
+
\sum_{m}
( f^{\dag}_{{L} m}  f^{\dag}_{{R} m} )
\left(
\begin{array}{cc}
    \widetilde{E}_{L} & \widetilde{V}_{\rm C} + \kappa\\
    \widetilde{V}_{\rm C} + \kappa & \widetilde{E}_{R}
\end{array}
\right)
\left(
\begin{array}{c}
 f_{{L} m} \\
 f_{{R} m}
\end{array}
\right) \nonumber\\
&&
+\frac{V}{\sqrt{N}}
\sum_{\alpha, k ,m }
b_{\alpha}
\left( 
c^{\dag}_{\alpha k m}f_{\alpha m}  + f_{\alpha m}^{\dag} c_{\alpha k m}
\right)
+
\sum_{\alpha = {\rm L, R}} \lambda_{\alpha} ( b_{\alpha}^{2} - 1 ) +
\frac{1}{J} \kappa^2,
\label{eq:HDiag}
\end{eqnarray}
where $\widetilde{E}_{\alpha} =  E_{\alpha} + \lambda_{\alpha}$ and
$\widetilde{V}_{\rm C} = \bar{b}_{L} \bar{b}_{R} V_{C}$
with $\bar{b}_{\alpha} = b_{\alpha} / \sqrt{N}$.

From this Hamiltonian,
we can calculate
the ground state energy.
Minimizing the ground state energy,
we determine five parameters $b_{\alpha},\;\lambda_{\alpha} \;
(\alpha={\rm L,R})$ and $\kappa$ self consistently.
(See Appendix A.)
In general, the self-consistent equations have
two solutions;
one is $\widetilde{\Delta}_{\alpha} = 0 \; (b_{\alpha}=0)$ and $\kappa = J/2$, and
the other is $\widetilde{\Delta}_{\alpha} \neq 0$.
The former means that
electrons in the dots are isolated from the leads and
make a spin-singlet state.
The latter corresponds to the coherent couplings between dots and
leads.
We determine the ground state by
comparing  the energies of these solutions.

\subsection{Conductance}
Next we derive the expression of the conductance $G$.
Current $I$ through the coupled dots
under a source-drain voltage $V_{\rm sd}$ is given by
a charge transfer through the central tunneling barrier
\begin{equation}
  I  =
  \frac{- i e V_{C}}{\hbar} \sum_{m}
 \left( \langle c_{{L}m}^{\dagger} c_{{R}m} \rangle -
 \langle c_{{R}m}^{\dagger} c_{{L}m} \rangle \right),
\label{eq:Current_Def}
\end{equation}
where $ \langle ... \rangle$ represents
the Keldysh-Green function.
By the mean-field approximations,
$\langle c_{{L}m}^{\dagger} c_{{R}m} \rangle =
b_{L} b_{R} \langle f_{{L}m}^{\dagger} f_{{R}m} \rangle$,
where
$\langle f_{{L}m}^{\dagger} f_{{R}m} \rangle$ can be calculated
using the Hamiltonian (\ref{eq:HDiag}). (See Appendix B.)

Evaluating the Keldysh-Green functions,
we obtain
the expression of the conductance $G \equiv \lim_{V_{\rm sd} \rightarrow
0} I/V_{\rm sd}$,
\begin{equation}
    G= \frac{2 e^{2}}{h} T(\omega=0),
\end{equation}
formally in the same form as the Landauer formulae.~\cite{Aono98}
The transmission probability $T(\omega)$ through the two dots is
\begin{equation}
T(\omega) = \frac{4 \widetilde{\Delta}_{L} \widetilde{\Delta}_{R} t^2 }
     { \left[ (\omega-\widetilde{E}_{L} - i \widetilde{\Delta}_{L})
      (\omega-\widetilde{E}_{R} - i \widetilde{\Delta}_{R}) - t^2 \right]
						\left[ (\omega-\widetilde{E}_{L} + i \widetilde{\Delta}_{L})
       (\omega-\widetilde{E}_{R} + i \widetilde{\Delta}_{R}) - t^2 \right]},
\label{eq:Transmission}
\end{equation}
with
$\widetilde{\Delta}_{\alpha} =
\bar{b}_{\alpha}^2
\Delta$ ($\alpha={L,R}$) and
$t = \widetilde{V}_C + \kappa$.
We choose $\omega=0$ at the Fermi level in the leads.

\section{Symmetric double dots without $J$}
\label{sec3}

We begin with a symmetric double dot system
$E_{L}=E_{R}$
in the absence of the spin-spin coupling $J=0$.
Because of the symmetry between the two dots
$\bar{b}_{L} = \bar{b}_{R} \equiv \bar{b}$,
$n_{L} = n_{R} \equiv n/2$,
$\lambda_{L} = \lambda_{R}$, and
$\widetilde{E}_{L} = \widetilde{E}_{R} \equiv \widetilde{E}$.
$\kappa=0$
due to the absence of $J$.
Then the expression of the transmission probability $T(\omega)$ in
Eq. (\ref{eq:Transmission}) is simplified to
\begin{equation}
T(\omega) =
\frac{4  \widetilde{\Delta}^2 \widetilde{V}_{\rm C}^2}
        {\left[ \left( \omega - (\widetilde{E} + \widetilde{V}_{\rm C})\right)^2 +
\widetilde{\Delta}^2
\right]
\left[ \left( \omega - (\widetilde{E} - \widetilde{V}_{\rm C}) \right)^2 + \widetilde{\Delta}^2
\right]},
\label{eq:conductance2}
\end{equation}
with
$\widetilde{\Delta} = \bar{b}^2 \Delta$ and
$\widetilde{V}_{\rm C} = \bar{b}^2 V_{C}$.
The parameter $\widetilde{V}_{\rm C}$
characterizes the effective dot-dot coupling.
Note that
$\widetilde{V}_{\rm C} / \widetilde{\Delta} =V_{C} / \Delta$.

The line shape of $T(\omega)$ is determined
by the parameter of $V_{C}/ \Delta$.~\cite{Aono98}
In Fig. \ref{fig2},
$T(\omega)$ is plotted as a function of $\omega/\widetilde{\Delta}$.
When $V_{C}/ \Delta < 1$,
$T(\omega)$ has a single peak at $\omega=\widetilde{E}$
[Fig.~\ref{fig2}(a)],
whereas
when  $V_{C}/ \Delta > 1$,
$T(\omega)$ has two maxima at
$\omega =\widetilde{E} \mp \sqrt{\widetilde{V}^2_{\rm C}-\widetilde{\Delta}^2}$
[Fig. \ref{fig2}(c)].
When $V_{C}/\Delta =1$,
$T(\omega)$ has a flat-topped single peak as shown in Fig. \ref{fig2}(b).

The dotted line in Fig. \ref{fig3}(a) represents
the conductance as a function of the gate voltage
when $V_C/\Delta < 1$.
For sufficiently low gate voltages,
the Fermi level in the leads is located at the peak of $T(\omega)$
[the case of $V_{g}/\Delta = -2$ is indicated by the arrow in Fig.~\ref{fig2}(a)].
In this case,
each dot accommodates one electron and
forms the Kondo resonant state with conduction electrons in a lead.
The electron transport is determined by the hopping between
the two Kondo resonant states.
Thus $G$ is proportional to $V_{C}^2$ and
independent of the gate voltage.
As the gate voltage increases,
the position of the Fermi level shifts downwards
[the cases of $V_{g}/\Delta = 0$ and $1$ in Fig. \ref{fig2}(a)].
The number of electrons in the dots $n$ decreases monotonically
from 2 to 0
(valence fluctuating regime).
As a result, $G$ decreases
with increasing the gate voltage.

The dotted line in Fig. \ref{fig4}(a) represents
the conductance as a function of the gate voltage
when $V_C/\Delta > 1$.
For low gate voltages,
the Fermi level in the leads is located at
the center of $T(\omega)$
[the case of $V_{g}/\Delta = -7$ in Fig. \ref{fig2}(c)],
and thus
the conductance $G$ is
considerably suppressed.
As the gate voltage increases,
the Fermi level shifts downwards
[the case of $V_{g}/\Delta = -4$ in Fig. \ref{fig2}(c)],
which increases $G$.
At a gate voltage of $V_{g} \approx 0$,
the Fermi level is just on the bonding peak and, thus,
the conductance $G$ has the maximum value of $2 e^2/h$.
On increasing the gate voltage further,
the Fermi level shifts downwards further
[the case of $V_{g}/\Delta = 5$ in Fig. \ref{fig2}(c)],
which leads to a decrease in $G$.

In summary,
when $V_{C}/\Delta < 1$,
the electron transport is characterized by
electronic states in each dot below the Kondo temperature
whereas,
when $V_{C}/\Delta > 1$,
it is characterized by
the formation of molecular levels of the Kondo resonant states.
In the recent work,~\cite{Aguado99}
Aguado and Langreth have argued that
these molecular levels are directly observable
in the differential conductance under
finite source-drain voltages.

We next mention $V_{g}$ dependence of
the peak width $\widetilde{\Delta}$ of the Kondo resonances.
When $V_{C}/\Delta \ll 1$,
\begin{equation}
 \widetilde{\Delta} =
       \cases{ T_{\rm K} &   ($V_{g} < 0$)  \cr
		             \widetilde{\Delta}_{\rm VF} \equiv \Delta/2
\left( \frac{1}{2} + \frac{2}{\pi}\arctan(V_{g}/\Delta)
\right) &  ($0 < V_{g}$ )
\cr },
\end{equation}
where $T_{\rm K}$ is the Kondo temperature for the coupled dots,
\begin{equation}
T_{\rm K} =
T_{K,1} \exp \left( \frac{V_{C}}{\Delta} \arctan \frac{V_{C}}{\Delta} \right)
     / \sqrt{ 1 + (V_{C} / \Delta)^{2}}.
\label{eq:Kondo_temp_strong}
\end{equation}
$T_{K,1}$ is given by Eq. (\ref{eq:Kondo_temperature}).
$T_{\rm K}$ increases exponentially with
$V_{g}$ or $V_{C}$.~\cite{Georges}
When $V_{C}/\Delta \gg 1$,
\begin{equation}
 \widetilde{\Delta} =
       \cases{
              T_{\rm K} &
                                              ($V_{g} < -V_{C}/2$)  \cr
     \frac{\Delta}{2 V_{C}} (V_{g} + V_{C}/2) &
                                              ($- V_{C}/2 < V_{g} <V_{C}/2$) \cr
		   \widetilde{\Delta}_{\rm VF} &
                                              ($V_{C}/2 < V_{g} $) \cr           
}.
\label{eq:K_temp_large}
\end{equation}
(See Appendix C.)

\section{Symmetric double dots with $J$}
\label{sec4}

Now
we discuss
a symmetric double dot system,
$E_{L}=E_{R}$,
in the presence of the spin-spin coupling $J$.
When the coupling $J$ is strong enough,
the Kondo coupling between a dot and lead is destroyed
and a spin-singlet state appears between the dots.
First we make a rough estimation of $J = J_c$,
where the crossover between the Kondo and spin-singlet states takes places.
The ground-state energy $\epsilon_{\rm gs}$ in Eq.~(\ref{eq:Egs}) is given by
\begin{equation}
\epsilon_{\rm gs} = V_g - \frac{2}{\pi}\widetilde{\Delta} - \frac{\kappa^2}{J}
           - \frac{2\widetilde{E}\widetilde{\Delta}}{\Delta},
\label{eq:Eg_J}
\end{equation}
where $\widetilde{\Delta}_{L,R} \equiv \widetilde{\Delta}$,
$\widetilde{E}_{L,R} \equiv \widetilde{E}$, and
$n_{L,R} \equiv n/2$.
Here we have used the self-consistent equations
in Appendix~A.
The last term is negligible when $n \sim 2$.
We consider two simple situations;
(i) Kondo state without spin-singlet ordering ($\kappa = 0$),
$\epsilon_{\rm gs} = V_g - \frac{2}{\pi}\widetilde{\Delta}$, and
(ii) spin-singlet state without the Kondo effect
($\widetilde{\Delta}=0$ and $\kappa = J/2$), $\epsilon_{\rm gs} = V_g - J/4$.
The crossover between (i) and (ii) takes place at
$J_{\rm c} / \widetilde{\Delta} = 8/\pi \cong 2.54$.
If $V_{C} \rightarrow 0$, $\widetilde{\Delta} \rightarrow T_{K,1}$
and thus
$J_{\rm c} / T_{K,1} = 2.54$.
This is consistent with the NRG result of the two impurity Kondo problem,
$J_{\rm c} / T_{K,1} \cong 2.2$.~\cite{Jones}
When $V_{C}/\Delta = 0.11$ and
$J_{\rm c}/\Delta =  2.1 \times 10^{-3} (U/\Delta = 21.2)$,
$J_{\rm c}/\Delta = 2.55 $ is in good agreement with
the recent NRG calculation of the coupled dot system \cite{Izumida99}
that gives
$J_{\rm c} / T_{K,1} = 2.65$.

In coupled quantum dots,
$J$ is given by $4 V_{C}^{2}/U$.
$U/\Delta = 4 \times 10^{2}$ is assumed in this paper.
We investigate the gate voltage dependence of $G$ with fixed $J$.
With decreasing $V_{g}$, the Kondo effect becomes weaker,
as indicated in Eq.~(\ref{eq:Kondo_temperature}),
and thus $J$ is relatively stronger compared with $\widetilde{\Delta}$.
At sufficiently low gate voltages, 
a spin-singlet state appears that destroys the Kondo effect.

\subsection{Transmission probability}
In the present case, the transmission probability $T(\omega)$ is given by
\begin{equation}
T(\omega) =
\frac{4  \widetilde{\Delta}^2 t^2}
        {\left[ \left( \omega - (\widetilde{E} + t)\right)^2 + \widetilde{\Delta}^2 \right]
\left[ \left( \omega - (\widetilde{E} - t) \right)^2 + \widetilde{\Delta}^2 \right]},
\label{eq:conductance3}
\end{equation}
where $t = \widetilde{V}_{\rm C} + \kappa$,
$\widetilde{V}_{\rm C} = \bar{b}^{2} V_{C}$,
and $\widetilde{\Delta} = \bar{b}^{2} \Delta$.
This result is the same as the one
without $J$ [Eq. (\ref{eq:conductance2})]
if
the dot-dot coupling $\widetilde{V}_{\rm C}$
is replaced by $t$.
Then
the line shape of $T(\omega)$ qualitatively depends on
whether $t/\widetilde{\Delta} = V_{C} / \Delta + \kappa/ \widetilde{\Delta}$
is larger than unity.

\subsection{Weak dot-dot coupling case}
First we discuss electron transport
when $V_{C} / \Delta < 1$.
The solid line in Fig.~\ref{fig3}(a) represents
$V_g$ dependence of $G$, 
while the solid and dotted lines
in Fig.~\ref{fig3}(b) represent $V_g$ dependence of $\kappa$ and
$\widetilde{\Delta}$, respectively
($V_C/\Delta = 0.3, J/\Delta = 9.0 \times 10^{-4}$).
For sufficiently low gate voltages,
$\kappa = J/2$ and $\widetilde{\Delta} = 0$,
indicating a complete spin-singlet state,
which results in $G=0$.
With increasing $V_g$,
the Kondo effect is stronger.
As a consequence,
$\widetilde{\Delta}$ becomes larger, whereas
$\kappa$ becomes smaller.
At high gate voltages,
$G$ is identical to the one without $J$ (dotted line).
Here the spin-spin coupling $J$ is weak enough compared with
the Kondo effect.
Between these two regions, $G$ has a sharp peak of $2 e^2/h$
in height.
At this peak,
both of the Kondo effect $\widetilde{\Delta}$ and
spin-singlet coupling $\kappa$ coexist
[see the inset in Fig. \ref{fig3}(b)].
This peak structure of $G$ is consistent with
the NRG calculations.~\cite{Izumida99}

The calculated results of the $G$-$V_{g}$ curve
can be understood
by Eq.~(\ref{eq:conductance3}).
At high gate voltages,
$\kappa/\widetilde{\Delta} \ll 1$ so that
$t/\widetilde{\Delta} = V_C/\Delta + \kappa/\widetilde{\Delta} \simeq V_C/\Delta$;
the plot of $G$ is similar to the one of $J=0$.
With decreasing $V_g$,
$\kappa / \widetilde{\Delta}$ increases and
thus $t/ \widetilde{\Delta}$ becomes larger.
On the right hand side of the sharp peak of $G$,
$t/\widetilde{\Delta} < 1$ and hence $T(\omega)$ has a single peak.
At $\kappa/\widetilde{\Delta} = 1 - V_C/\Delta$,
$t/\widetilde{\Delta} =1$ and $T(\omega = 0) = 1$,
resulting in $G=2 e^2/h$.
With decreasing $V_g$ further,
split peaks of $T(\omega)$ suppress $G$.
At a certain gate voltage,
the Kondo states disappear and a spin-singlet state appears
($\kappa = J/2$ and $\widetilde{\Delta}=0$).

From the results given above,
 the origin of the peak of $G$ is explained as follows.
The Kondo resonant state is formed by
the coherent coupling between the dot and the lead
while the spin-singlet state
is formed by the coherent coupling
between the two dots.
When these two states coexist,
the left and right leads are connected
via a coherent channel.
Electrons transfer through this channel
and as a result, $T(\omega=0) = 1$.

We find that
the width of the sharp peak is roughly given by $V_C$.
This implies
that
the crossover region between
the spin-singlet state and the Kondo state
becomes wider as $V_{C}$ increases.
It is consistent with the results by
Sakai {\it et al.} \cite{Sakai90,Sakai92}
for the two-impurity Kondo problem.
(See also Refs. \onlinecite{Izumida99,Georges}.)

\subsection{Strong dot-dot coupling case}
Next we discuss electron transport when
$V_{C} / \Delta > 1$.
The solid line in Fig.~\ref{fig4}(a) represents
$V_g$ dependence of $G$;
the solid and dotted lines
in Fig.~\ref{fig4}(b) represent $V_g$ dependence of $\kappa$ and
$\widetilde{\Delta}$, respectively
($V_C/\Delta = 10.0, J/\Delta = 1.0$).
For this case,
the gate voltage dependence of $G$ is fairly robust
against $J$ except that the region of $G=0$ for low gate voltages.
When $G=0$,
$\kappa = J/2$ and $\widetilde{\Delta} = 0$
(spin-singlet state in the dots)
as in the case of $V_C/\Delta < 1$.
Both of $\kappa$ and $\widetilde{\Delta}$ change gradually
against the gate voltage
(there is a simple relation of
$2\kappa/J = 1/2 - \widetilde{\Delta}/\Delta$, see Appendix C).
These results contrast to the ones for
the weak dot-dot coupling case.

The robustness of $G$ against $J$ is explained as follows.
At the top of the broad peak of $G$,
$n \simeq 1$ as seen in the previous section.
On the right-hand side of the peak,
$n < 1$ and
spin-spin coupling $J$ is not relevant.
On the left-hand side of the peak,
$n>1$ and
the antiferromagnetic interaction becomes more effective
with decreasing $V_g$.
However,
the transition from the Kondo states to the spin-singlet state
takes place at quite a low gate voltage
because
the Kondo temperature $T_{\rm K}$ [Eq. (\ref{eq:Kondo_temp_strong})]
is much larger than $T_{\rm K,1}$ for strong dot-dot coupling.
In addition,
$t/\widetilde{\Delta}$ is always larger than unity
when $V_C/\Delta > 1$.
Thus
the spin coupling $J$ does not introduce
qualitatively new features in
expression (\ref{eq:conductance3}).

\subsection{Intermediate region of $V_{C}/\Delta$}
Finally, we show
$V_g$ dependence of $G$
by changing the magnitude of $V_{C}/\Delta$
in Fig.~\ref{fig:crossover}.
As $V_{C}$ increases,
(i) the height of the plateau increases,
(ii) the sharp peak of $G$ broadens,
and
(iii) the position of the sharp peak of $G$ shifts to larger $V_g$.
When $V_{C}/\Delta >1$,
the peak position is fixed at $V_{g} \sim 0$.
At the peak of $G$, $n=2$ when $V_{C}/\Delta \ll 1$ while
$n =1 $ when $V_{C}/\Delta \gg 1$;
$n$ decreases gradually from 2 to 1 as $V_{C}$ increases.
In conclusion,
electron transport and electronic states in the dots
change continuously from the weak to strong dot-dot coupling regions.

\section{Asymmetric Double Quantum Dots}
\label{sec5}

In this section,
we discuss the effects of the energy difference $\Delta E = E_{L} - E_{R}$
between the dots.
First we describe the case of $V_C/ \Delta < 1$.
In Fig. \ref{fig:weak_asym},
the gate voltage dependence of $G$ is shown for
$\Delta E=0$ (thin dotted line)
and $\Delta E \neq 0$ (thick dotted line)
in the absence of $J$.
These plots show that
$\Delta E$ has no effect on the height of $G$
when the gate voltage is low enough
and both of the dots accommodate one electron.
This is because
the dot-lead coupling makes the Kondo resonant state
at the Fermi level in the leads,
irrespective of the position of the dot levels.
The single peak of the transmission probability $T(\omega)$ is located at
the Fermi level, as discussed in Fig. \ref{fig2}(a).
As a result, $G$ is insensitive to $\Delta E$.
In the presence of $J$,
the gate voltage dependence of $G$ is shown for
$\Delta E=0$ (thin solid line)
and $\Delta E \neq 0$ (thick solid line).
The sharp peak of $G$ appears, but
the peak height is less than $2e^2/h$.

Next, we describe the case of $V_{C}/\Delta > 1$.
In Fig. \ref{fig:strong_asym},
the gate voltage dependence of $G$ is shown for
$\Delta E=0$ (thin dotted line)
and $\Delta E \neq 0$ (thick dotted line) 
in the absence of $J$.
For low gate voltages,
$n_L=n_R=1$ and
$G$ is independent of $\Delta E$
as in the case of $V_C / \Delta < 1$.
In the valence fluctuating regime,
where $n$ decreases as $V_g$ increases,
$n_L \neq n_R$,
and hence, electron transport depends on $\Delta E$ prominently.
The broad peak of $G$ splits into two peaks.
They are less than $2 e^2/h$ in height.
At the left-hand peak,
$n_L$ decreases from unity  to zero
with increasing $V_g$ while $n_R \simeq 1$.
At the right-hand peak,
$n_R$ decreases from unity to zero
while  $n_L \simeq 0$.
These peaks are understood by the Kondo resonance
between a molecular orbital in the dimer and leads. One
molecular orbital has the amplitude mainly in the left dot, and
the other in the right dot. The peak heights of $G$ are determined by the
amplitude of the orbitals in the other dot.
Note that
the gate voltage dependence of $G$ is fairly robust
against $J$
as in the case of $\Delta E =0$
(compare the thick dotted and thick solid lines in Fig. \ref{fig:strong_asym}).

Finally, we show the intermediate region of $V_C/\Delta$ in
Fig. \ref{fig:crossover_asym}.
When $V_C / \Delta \ll 1$,
$G$ has a sharp peak.
As $V_C/\Delta$ increases,
the sharp peak broadens and
a weak peak develops at $V_{g} \sim \Delta E/2$.
The width of the weak peak is $\sim \Delta$.
The line shapes of these two peaks become similar to each other.
Electron transport
changes continuously from the weak to strong dot-dot coupling regions,
as discussed in Fig. \ref{fig:crossover} in the case of $\Delta E = 0$.

\section{Conclusions and Discussion}
\label{sec6}

We have investigated the conductance
through coupled quantum dots in series,
as a function of the gate voltage,
below the Kondo temperature.
Electron transport is characterized by
the ratio of $V_{C} / \Delta$.
When $V_{C} / \Delta < 1$,
a plateau of $G$ appears
because each dot forms the Kondo resonant state with
conduction electrons in a lead and $G$ is determined by the hopping between
the Kondo states.
For low gate voltages, they are replaced by
the spin-singlet state and $G=0$.
A sharp peak of $G$  appears with the height of $2 e^2/h$,
which is due to
a coherent electron transport brought by
the cooperation between the Kondo and spin-singlet states.
When the energy levels are different in the dots,
the sharp peak survives, but the peak height is less than $2 e^2/h$.
When $V_{C} / \Delta > 1$,
 the molecular levels between the Kondo resonant states appear.
The gate voltage dependence of $G$ has a broad peak.
The antiferromagnetic spin coupling is ineffective, compared
with the case of  $V_{C}/\Delta < 1$,
because (i) the Kondo temperature of a coupled dot is large
and (ii) the transmission probability
always has two peaks.
When the energy levels are different in the dots,
the conductance $G$ has two peaks,
reflecting asymmetric molecular levels.

We should mention
the validity of the mean-field approximation.
First,
the mean-field theory predicts the first-order transition
between the Kondo and spin-singlet states
when  $V_{C}/\Delta <  1/\pi$;~\cite{Jones2}
it is an artifact of the mean field theory, because
the NRG results show that when $V_{C} \neq 0$,
the transition is always smooth.~\cite{Sakai90,Sakai92}
For the present problem of coupled dot systems,
this artifact appears on
the left-hand side of the sharp peak of $G$;
a discontinuous jump of $G$ is seen for $V_C / \Delta = 0.2$ in Fig. \ref{fig:crossover}.
Second,
in the valence fluctuating regime,
the fluctuations around the mean field of the slave boson are not negligible.
Hence, our calculations are not enough for quantitative discussion.

In real coupled quantum dots,
the intersite Coulomb energy, $U_{\rm int} n_{L} n_{R}$,
is present.
Then the ground-state energy $E_{\rm g, int}$ is given by
\begin{eqnarray}
   E_{\rm g, int}  & = & E_{\rm gs} +
   U_{\rm int} ( 1- b_{L}^{2})  ( 1- b_{R}^{2}),
    \label{eq:Egs_intersite}
\end{eqnarray}
in the mean-field approximation.
Equation (\ref{eq:Egs_intersite}) indicates that
the intersite Coulomb energy raises the dot levels $E_L$ and $E_R$
effectively.
As a result,
the valence fluctuating regime is extended and
the Kondo regime shifts to lower $V_g$.
Note that
we have assumed that the on-site Coulomb interaction $U$ is infinite,
and thus, $U_{\rm int} \ll U$.

When $U_{\rm int}$ is large enough,
the coupled dots cannot accommodate  more than one electron.
This situation has been investigated by
Pohjola {\it et al.}\cite{Pohjola}.
In this case,
the coupled dots is regarded as
a single quantum dot with two levels
when the dot-dot tunneling is large.
Pohjola {\it et al.} have found a splitting of the Kondo resonance due to
the level difference $\Delta E$ between the two dots.
The one peak is on the Fermi level $E_F$ and
the other is located at $E_F+\Delta E$.
This is in contrast to the present case in which
the split peaks are located below and above $E_F$.

Finally, we comment on an asymmetric dot-lead coupling case,
$\Delta_{L} \neq \Delta_{R}$.
This asymmetry introduces the following two effects.
First, the maximum of the conductance is less than $2 e^2/h$
by a geometrical asymmetry
$ 4 \Delta_{L} \Delta_{R} / (\Delta_{L} + \Delta_{R})^2$.~\cite{Glazman}
Second,
the difference of dot-lead coupling $\delta \Delta \equiv (\Delta_{L} - \Delta_{R})/2$
introduces a difference in the number of electrons between the dots,
resulting in $\widetilde{E}_{L} \neq \widetilde{E}_{R}$.
Electronic states are, therefore, asymmetric;
the situation is similar to the case of $E_L \neq E_R$ discussed in
Sec.~\ref{sec5}.
[Equation (\ref{eq:xi_def}) in Appendix A is modified by replacing
$\widetilde{\Delta}_0 \rightarrow
\widetilde{\Delta}_0 + \bar{b}_1^2 \; \delta \Delta$ and
$\widetilde{\Delta}_1 \rightarrow
\widetilde{\Delta}_1 + \bar{b}_0^2 \;\delta \Delta$.
The latter modification means that
$n_L \neq n_R$ when $\delta \Delta \neq 0$.]

\section*{ACKNOWLEDGMENTS}
We acknowledge  Yu.~V.~Nazarov for fruitful comments.
T.A. is supported by the
Special Postdoctoral Researchers Program of RIKEN.
Numerical calculations were performed on the workstation in
the Computer Information Center, RIKEN.

\appendix

\section{Derivation of the ground state energy}
First, we review a single impurity model with the impurity level
$E$.~\cite{Coleman87,Newns88,HewsonBook}
We start from the following mean-field Hamiltonian ${\cal
H}$
\begin{eqnarray}
{\cal H} &=&
\sum_{k, \; m}
 E(k) c^{\dag}_{k m}c_{k m}
+
\sum_{m}
 \widetilde{E}  f^{\dag}_{m}  f_{m}
+\frac{V b}{\sqrt{N}}
\sum_{k ,m }
\left( 
c^{\dag}_{ k m}f_{ m}  + f_{ m}^{\dag} c_{ k m}
\right)
+
\lambda ( b^{2} - 1 )
\label{eq:HDiag_single}
\end{eqnarray}
with $\widetilde{E} =  E+ \lambda$ and $b$ is the mean field value of
the slave boson.

The free energy $F$ of the system is given by
\begin{equation}
    F = - \frac{2}{\beta} \int {\rm d} \omega
    \ln (1 + e^{- \beta  \omega}) \bar{\rho}(\omega)
+
\lambda ( b^{2} - 1 ),
    \label{eq:free_energy}
\end{equation}
where
$\bar{\rho}(\omega)$ is the density of states,
which is given by the retarded Green function $G_r(\omega)$ of
electrons in the dot
\begin{equation}
    \bar{\rho}(\omega) = \frac{\rm Im}{\pi} G_r(\omega)
     = \frac{\rm Im}{\pi} \frac{1}{\omega - \widetilde{E} - \Sigma(\omega)},
    \label{eq:dos}
\end{equation}
with the self-energy $\Sigma(\omega)$
\begin{equation}
\Sigma(\omega) = \frac{b^2 V^2}{N} \sum_k \frac{1}{\omega - E(k) + i \eta}
               =  - i\pi \frac{b^2 V^2}{N} \rho(\omega) \equiv -i \widetilde{\Delta}
\Phi(\omega).
\label{eq:imag}
\end{equation}
Here $\rho(\omega)$ is the density of states of conduction electrons in a lead,
$\rho = \rho(\omega=0)$, $\Phi(\omega) = \rho(\omega)/\rho$, and
$\widetilde{\Delta} =  \pi \rho b^2 V^2/N $.
To derive Eq. (\ref{eq:imag}),
we have assumed that
$\rho(\omega)$ varies smoothly when compared with
$\widetilde{\Delta}$.

By the substitution of Eqs. (\ref{eq:dos}) and (\ref{eq:imag})
into Eq. (\ref{eq:free_energy}) and integration by parts,
the free energy (\ref{eq:free_energy}) becomes
\begin{equation}
 F = \frac{N}{\pi} {\rm Im} \int_{-\infty}^{\infty} {\rm d}\omega
 f(\omega) \Phi(\omega) \ln \left( \omega - \xi \right) +
\lambda ( b^{2} - 1 ),
\end{equation}
with $ \xi = \widetilde{E} + i \widetilde{\Delta}$.

If the leads has a wide flat band of width $2D$,
$\Phi(\omega) = 1 (-D < \omega < D)$,
the free energy is
\begin{equation}
 F = \frac{N}{\pi} {\rm Im} \int_{-D}^{D} {\rm d}\omega
 f(\omega) \ln \left( \omega - \xi \right)
+
\lambda ( b^{2} - 1 ).
\label{eq:free_energy_wide_band}
\end{equation}
At zero temperature,
this yields
\begin{equation}
F = \frac{N}{\pi} {\rm Im} \left\{ \xi \left( \ln \left(\frac{\xi}{D} \right) -1
\right)
\right\}+
\lambda ( b^{2} - 1 ).
\end{equation}

An extension to the coupled dot system is straightforward.
The free energy is
\begin{equation}
 F = \frac{N}{\pi} {\rm Im} \sum_{P=\pm} \int_{-D}^{D} {\rm d}\omega
 f(\omega) \ln \left( \omega - \xi_{P} \right)
+
\sum_{\alpha = {\rm L, R}} \lambda_{\alpha} ( b_{\alpha}^{2} - 1 ) +
\frac{N}{J} \kappa^2,
\label{eq:F_final}
\end{equation}
with
\begin{equation}
\xi_{P}= \widetilde{E}_{0} + i \widetilde{\Delta}_{0} +
P \sqrt{(\widetilde{E}_{1}+ i \widetilde{\Delta}_{1})^2 + t^2}.
\label{eq:xi_def}
\end{equation}
Here
$\widetilde{E}_{0,1} = (\widetilde{E}_{L} \pm \widetilde{E}_{R})/2$,
$\widetilde{\Delta}_{0,1} = \bar{b}_{0,1}^{2} \Delta$
with
$\bar{b}_{0,1}^{2} =
(\bar{b}_{L}^{2} \pm \bar{b}_{R}^{2})/2$,
and
$t = V_{C} \sqrt{\bar{b}_{0}^{4}- \bar{b}_{1}^{4}} + \kappa$.
$\xi_{P}$ are the eigenvalues of the matrix given by
\begin{equation}
\left(
\begin{array}{cc}
    \widetilde{E}_{L} + i \widetilde{\Delta}_{L}&
    \widetilde{V}_{\rm C} +\kappa\\
    \widetilde{V}_{\rm C} + \kappa &
     \widetilde{E}_{R}+ i \widetilde{\Delta}_{R}
\end{array}
\right).
\end{equation}

Then the ground-state energy $E_{\rm gs} = N \epsilon_{\rm gs}$[$= F(T=0)$] is
written as
\begin{equation}
\epsilon_{{\rm gs}} =
2 ( \widetilde{E}_{0} - E^{*}) ( \bar{b}_{0}^{2} - 1/2)
+ 2(\widetilde{E}_{1} - \Delta E) \; \bar{b}_{1}^{2} + \frac{1}{J}
\kappa^{2}  + \frac{1}{\pi} \sum_{P=\pm} {\rm Im} \;
\left[ \xi_{P} \left( \ln \frac{\pi \xi_{P}}{\Delta} - 1 \right) \right],
\label{eq:Egs}
\end{equation}
where
$E^{*} =  V_{g} + \Delta/\pi
\ln \left( D \pi/\Delta \right)$
with the bandwidth $D$ in the leads.~\cite{Haldane78}
In experiments,
$E^{*}$ can be controlled by the gate voltage $V_{g}$
(we redefine $E^{*}$ as $V_{g}$).

Minimizing the ground-state energy (\ref{eq:Egs}),
we determine five parameters, $b_{\alpha},\;\lambda_{\alpha} \;
(\alpha={\rm L,R})$, and $\kappa$, self-consistently.
The self-consistent equations are
\begin{eqnarray}
    & & 2 \left( \widetilde{E}_{0} - V_{g} \right)
    + \frac{1}{\pi} \sum_{P} {\rm Im}
    \left[
    \left(
    i \Delta +
    \frac{P \; \bar{b}_{0}^{2} \; t \; V_{C} }
    {\sqrt{ \phantom{\bigl[} \bar{b}_{0}^{4}-\bar{b}_{1}^{4}}
     \sqrt{(\widetilde{E}_{1}+ i \widetilde{\Delta}_{1})^2 + t^2}}
    \right)
    \ln \frac{\pi \xi_{P}}{\Delta}
    \right] = 0,
    \label{eq:SCE1}\\
    & & 2 \left( \widetilde{E}_{1} - \frac{\Delta E}{2} \right) 
    + \frac{1}{\pi} \sum_P {\rm Im}
    \left[
    \frac{i \widetilde{E}_{1} \Delta  -
    \bar{b}_{0}^{2} \left(
    \Delta^{2} + V_{C}^{2} +
    \kappa V_{C}/\sqrt{ \phantom{\bigl[} \bar{b}_{0}^{4}-\bar{b}_{1}^{4}}
    \right)}
    {\sqrt{(\widetilde{E}_{1}+ i \widetilde{\Delta}_{1})^2 + t^2}}
    P \ln \frac{\pi \xi_{P}}{\Delta}
    \right] = 0,
    \label{eq:SCE2}\\
    & & 2 \left( \bar{b}_{0}^{2} -1/2 \right) 
    + \frac{1}{\pi} \sum_{P} {\rm Im} \left[ \ln \frac{\pi \xi_P}{\Delta} \right] = 0,
    \label{eq:SCE3}\\
    & & 2 \bar{b}_{1}^{2}
    + \frac{1}{\pi} \sum_{P} {\rm Im}
    \left[ \frac{(\widetilde{E}_{1} + i \widetilde{\Delta}_{1})}
    {\sqrt{(\widetilde{E}_{1}+ i\widetilde{\Delta}_{1})^2 + t^2}}
    P \ln \frac{\pi \xi_P}{\Delta} \right] = 0,
     \label{eq:SCE4}\\
    & & \frac{2 \kappa}{J} + \frac{1}{\pi} \sum_P {\rm Im}
    \left[ \frac{t}{\sqrt{(\widetilde{E}_{1}+ i \widetilde{\Delta}_{1})^2 + t^2}}
    P \ln \frac{\pi \xi_P}{\Delta} \right] = 0.
     \label{eq:SCE5}
\end{eqnarray}

\section{Expression of current}
In this appendix,
we derive the expression of the current
using the real-time functional integral method.
The alternative derivation is given in Ref. \onlinecite{Kawamura},
which is based on the equation of motions for
the Green functions.
We take $\hbar = 1$ in this appendix.

All of the physical quantities can be obtained in principle by
calculating the time evolution of the density matrix $\rho(t)$
\begin{equation}
 \rho(t)  =  U^{\dagger}(t) \rho_{\rm init.} U(t),
\label{eq:density_mat_def}
\end{equation}
where $\rho_{\rm init.}$ is the initial density matrix and $U(t)$ is the time evolution kernel of
the Hamiltonian $H$
\begin{equation}
 U(t) = T \exp \left( -i H(t)\right),
\end{equation}
using the time ordering product $T$.

Corresponding to Eq. (\ref{eq:density_mat_def}),
we introduce the generating functional
$W[{\bbox{j}}_{1},\bbox{j}_{2}]$,~\cite{schwinger,fukuda-nomoto}
\begin{equation}
  \exp \left( W \left[{\bbox{j}}_{1},\bbox{j}_{2} \right] \right) =
    {\rm Tr}\; U^{j_{1} \; \dagger}(t) \rho_{\rm init.} U^{j_{2}}(t),
\label{eq:generating_func_def}
\end{equation}
where $U^{j}(t)$ is the time evolution kernel of
the Hamiltonian
$H^{j} = H - \bar{{\bbox j}} \cdot {\bbox f} - \bar{{\bbox f}} \cdot {\bbox j} $
with
$\bar{{\bbox j}} = ( \bar{j}_{{L} m}, \bar{j}_{{R} m} )$,
$\bar{{\bbox f}} = ( \bar{f}_{{L} m}, \bar{f}_{{R} m} )$ and
their conjugates ${\bbox j}$ and ${\bbox f}$.
Equation (\ref{eq:generating_func_def}) is evaluated using
the real-time functional integral method,
which is
the functional integral on the Keldysh contour;~\cite{keldysh,chen-ting-1,meir-1,meir-2}
the suffix ``1" and ``2" in $j$ correspond to the upper and lower branches of
the Keldysh contour, respectively.

For the  Hamiltonian (\ref{eq:HDiag}),
Eq. (\ref{eq:generating_func_def}) is given by
the following functional integral
on the degrees of freedom
of electrons in the dots;~\cite{omitt}
\begin{equation}
    \exp \left( W[J] \right) =  \int {\cal D}{\bbox f} {\cal D}\bar{\bbox f}
    \; \exp \left( \sum_{m=\pm1/2} \int {\rm d}\omega\;
    \bar{{\bbox F}}\; G^{-1} \; {\bbox F}-
    {\bbox J} \cdot \bar{{\bbox F}} - \bar{{\bbox F}} \cdot {\bbox J}
    \right)
      \label{eq:E}
\end{equation}
$\bar{{\bbox F}} = \left(
    \bar{f}_{+;{L}m}(\omega),\; \bar{f}_{-;{L}m}(\omega),\;
    \bar{f}_{+;{R}m}(\omega),\; \bar{f}_{-;{R}m}(\omega)
    \right)$ and
$\bar{{\bbox J}} = \left(
    \bar{j}_{-;{L}m}(\omega),\; \bar{j}_{+;{L}m}(\omega),\;
    \bar{j}_{-;{R}m}(\omega),\; \bar{j}_{+;{R}m}(\omega)
    \right)$
with
$f_{+} = (f_1 + f_2)/2$, $f_{-} = f_1-f_2$,
$j_{+} = (j_1 + j_2)/2$, $j_{-} = j_1-j_2$.
The matrix $G^{-1}$ is the inverse of the Green functions:
\begin{equation}
     G^{-1} (\omega ) =
     \left(
    \begin{array}{cccc}
     0 & G^{-1}_{\rm r, L}(\omega) & 0 & t\\
     G^{-1}_{\rm a, L}(\omega) & \widetilde{\Delta}_{L} f_{L}(\omega) & t & 0\\
     0 & t & 0 & G^{-1}_{\rm r, R}(\omega)\\
     t & 0 & G^{-1}_{\rm a, R}(\omega) & \widetilde{\Delta}_{R} f_{R}(\omega)
   \end{array}
 \right),
\end{equation}
with the Fermi function $f_{\alpha}(\omega)$ of the electrodes with
the chemical potential $\mu_{\alpha}$ and
the retarded and advanced Green function
$G^{-1}_{{\rm r,a},\; \alpha} \; (\alpha = {\rm L,\;R})$ of electrons in the dots:
\begin{eqnarray}
    G^{-1}_{\rm r, \alpha}(\omega) & = &
    \omega + \widetilde{E}_{\alpha} +  i \widetilde{\Delta}_{\alpha},
      \\
     G^{-1}_{\rm a, \alpha}(\omega) & = &
     \omega + \widetilde{E}_{\alpha} - i \widetilde{\Delta}_{\alpha}.
\end{eqnarray}
After integrating over the degrees of freedom of electron in the dots,
the generating functional $W$ is finally given
by $G(\omega)$,
\begin{equation}
    W = - \sum_{m = \pm 1/2}\int d \omega\;
    \bar{{\bbox J}}\; G (\omega)\; {\bbox J}.
\end{equation}

The current $I$ (\ref{eq:Current_Def}) is rewritten as
\begin{eqnarray}
    I &=& - i e \widetilde{V}_{\rm C} \sum_m
     \int d \omega
    \left(
    \langle \bar{f}_{+;{L}m} f_{+;{R}m}(\omega) \rangle -
    \langle \bar{f}_{+;{R}m} f_{+;{L}m}(\omega) \rangle
    \right).
 \end{eqnarray}
This is represented using the generating functional $W$
 \begin{eqnarray}
     I	& =&  - i e \widetilde{V}_{\rm C} \sum_m \int d \omega
												\left(
																			\frac{\delta}{\delta J_{-;{L}m}}
                   \frac{\delta}{\delta \bar{J}_{-;{R}m}}
																		-
                   \frac{\delta}{\delta J_{-;{R}m}}
                   \frac{\delta}{\delta \bar{J}_{-;{L}m}}
            \right)
	     \exp ( W[J_+,J_-])\\
     &=&  - 2 i e \widetilde{V}_{\rm C} \int d \omega
           \left[ G_{1,3}(\omega) - G_{3,1}(\omega) \right],
\end{eqnarray}
where $G_{n,m}(\omega)$ is the $(n,m)$ component of $G$.
$I$ is finally given by
\begin{equation}
   I  =  2e \int  d \omega \; T(\omega)
\left[ f_{L}(\omega) - f_{R}(\omega) \right]
\label{eq:I}
\end{equation}
with
$T(\omega)$ in Eq. (\ref{eq:Transmission}).

To derive Eq. (\ref{eq:I}),
we have assumed $\widetilde{\Delta}_{\alpha}$, $\widetilde{E}_{\alpha}$, and $\kappa$
are determined by the self-consistent equations for the thermal equilibrium case
$\mu_{L}=\mu_{R}$.
Thus, the present calculation is valid only in the limit of $\mu_{L} \rightarrow \mu_{R}$.
Extensions to the case of finite $V_{\rm sd}$ is possible within the framework of
the present method.
See Refs. \onlinecite{Ivanov,Aguado99}.

\section{Kondo temperature when $V_{C}/\Delta \gg 1$}
In this appendix,
we derive the gate voltage dependence of $\widetilde{\Delta}$ when $J=0$.
The self-consistent equations are given by
\begin{eqnarray}
    (\bar{b}_{0}^{2} - 1/2) +
    \frac{1}{2\pi} \sum_{P} {\rm Im}
    \left[ \ln \left( \frac{\pi \xi_{P}}{\Delta} \right)
\right] & = & 0,
    \label{eq:SCE1_Jzero}  \\
    (\widetilde{E} - V_{g}) + \frac{1}{2\pi} \sum_{P} {\rm Im}
    \left[ \left( P V_{C} +  i \Delta \right )
\ln \left( \frac{\pi \xi_{P}}{\Delta} \right)
\right] & = & 0,    
    \label{eq:SCE2_Jzero}
\end{eqnarray}
where $\xi_{P} = \widetilde{E} + P \widetilde{V}_{\rm C} + i \widetilde{\Delta}$
with $\widetilde{V}_C = \bar{b}_0^2 V_C$ and $\widetilde{\Delta} = \bar{b}_0^2 \Delta$.

When $ V_{g} < - V_{C}/2$,
the Kondo state is formed in each dot and lead
and thus
$\widetilde{E} = 0$.
Then Eq. (\ref{eq:SCE2_Jzero}) is reduced to
\begin{eqnarray}
- V_{g} - \frac{V_{C}}{\pi} \arctan ( V_{C} / \Delta)  +\frac{\Delta}{\pi}
     \ln \left( \frac{\pi \sqrt{\widetilde{\Delta}^2 + \widetilde{V}_{\rm C}^2}}{\Delta} \right)
&=& 0,
\end{eqnarray}
resulting in the first line of Eq. (\ref{eq:K_temp_large}).

When $ -V_{C}/2  < V_{g} < V_{C}/2 $,
the broad peak of $G$ appears.
For this case,
the bonding level is located near the Fermi level in the leads and
the antibonding level is far away from the Fermi level.
Then we can use the following two conditions:
(i) $\widetilde{E} \cong \widetilde{V}_{\rm C}$ ( ${\rm Re} \xi_{-} \cong 0 $)
and (ii)
$
\sum_{P} {\rm Im} \left[  \ln \left( \frac{\pi \xi_{P}}{\Delta} \right) \right]
 \cong
 -\sum_{P} {\rm Im} \left[ P \ln \left(\frac{\pi \xi_{P}}{\Delta} \right) \right]
\cong
 {\rm Im} \left[ \ln \left( \frac{\pi \xi_{-}}{\Delta} \right) \right]
$.
Then substituting Eq. (\ref{eq:SCE1_Jzero}) into Eq. (\ref{eq:SCE2_Jzero})
yields
\begin{equation}
2 \widetilde{V}_{\rm C} - \left( V_{g} + \frac{1}{2} V_{C} \right)  + \frac{\Delta}{2\pi}
\sum_{P} {\rm Re}
    \left[ \ln \left( \frac{i \pi \widetilde{\Delta}}{\Delta} \right) +
           \ln \left( \frac{\pi \left( 2 \widetilde{V}_{\rm C} + i\widetilde{\Delta} \right)
}{\Delta}
\right)
\right] =  0.
\end{equation}
The logarithmic terms are negligible
because
$n \cong  1$ so that
$\widetilde{\Delta} \sim \Delta/2$ and $\widetilde{V}_{\rm C} \sim  V_{C}/2$.
Thus the gate voltage dependence of $\widetilde{\Delta}$ reduces to
the second line in
Eq. (\ref{eq:K_temp_large}).

When $V_{C}/2 < V_{g}$ where $n < 1$,
the coupled dot system is essentially the same as
the noninteracting system;
$n$ is determined by the position of the energy level:
\begin{equation}
  n  = 1- \frac{2}{\pi} \arctan \left( V_{g} / \Delta \right).
\end{equation}
This result yields the third line of Eq. (\ref{eq:K_temp_large}),
because $\widetilde{\Delta}  = \Delta(1 -n/2)/2$.

Finally, we mention the relation between $\kappa$ and $\widetilde{\Delta}$
when $J \neq 0$ (Sec. \ref{sec5}).
The self-consistent equation for $\kappa$, Eq. (\ref{eq:SCE5}),
\begin{equation}
\frac{2\kappa}{J} + \frac{1}{\pi} \sum_{P} {\rm Im} \;
     \left[ P \ln \left( \frac{\pi \xi_{P}}{\Delta} \right) \right]  =  0,
   \label{eq:SCE3_Jzero}
\end{equation}
is simplified for $ -V_{C}/2 <  V_{g} <  V_{C} /2$.
Using the conditions (i) and (ii),
we obtain
\begin{equation}
\frac{2\kappa}{J} +
\frac{\widetilde{\Delta}}{\Delta} - 1/2 = 0.
\end{equation}

\newpage

\begin{figure}[hbt]
    \centering
    \caption{%
    A quantum dot dimer connected in series.}
    \label{fig1}
\end{figure}
 
\begin{figure}[hbt]
    \centering
    \caption{%
The transmission probability $T(\omega)$ as a function of
$(\omega-\widetilde{E})/\widetilde{\Delta}$ with
 the energy of incident electrons $\omega$,
the level position $\widetilde{E}$ and width $\widetilde{\Delta}$ of the Kondo resonance.
$E_{L}=E_{R}$ and $J=0$.
(a) $V_{C}/\Delta=$
 0.3, (b) 1.0, and (c) 10.0.
Arrows indicate the positions of the Fermi level in the leads for
several values of the gate voltage $V_{g}/\Delta$.}
  \label{fig2}
\end{figure}

  \begin{figure}[hbt]
  \centering
  \caption{(a) The conductance $G$ versus the gate voltage
when $V_{C}/ \Delta = 0.3$. $E_{L} = E_{R}$.
The dotted and solid lines represent the cases of $J=0$ and
$J / \Delta = 9.0 \times10^{-4}$ ($U / \Delta = 4 \times 10^{2}$),
respectively.
(b) The Kondo resonance width $\widetilde{\Delta}$ (dotted line) and
the spin-singlet order parameter $\kappa$ (solid line) versus
the gate voltage for $J / \Delta = 9.0 \times10^{-4}$.
The inset shows $G$ (broken line), $\widetilde{\Delta}$ (dotted line), and $\kappa$
(solid line) as functions of the gate voltage, near the sharp peak of $G$.}
  \label{fig3}
  \end{figure}

  \begin{figure}[hbt]
  \centering
  \caption{%
(a) The conductance $G$
versus the gate voltage when $V_{C}/ \Delta = 10.0$. $E_{L}=E_{R}$.
The dotted and solid lines represent the cases of
$J=0$ and $J/\Delta = 1.0\; (U/\Delta = 4 \times 10^{2})$, respectively.
(b) The Kondo resonance width $\widetilde{\Delta}$ (dotted line) and
the spin-singlet order parameter $\kappa$ (solid line) versus
the gate voltage for $J/\Delta = 1.0$.
}
  \label{fig4}
  \end{figure}

  \begin{figure}[hbt]
  \centering
  \caption{%
 The conductance $G$ versus the gate voltage
for several values of $V_{C}/\Delta$ from 0.2 to 3.0. $E_{L}=E_{R}$.
All the curves are vertically offset for clarity.}
\label{fig:crossover}
\end{figure}

\begin{figure}[hbt]
  \centering
  \caption{%
The conductance $G$ versus the gate voltage when
$V_{C}/\Delta = 0.3$.
For $J=0$, $\Delta E/\Delta = 0.0$ (thin dotted line) and 2.0
(thick dotted line).
For $J / \Delta = 9.0 \times10^{-4}$ ($U / \Delta = 4 \times 10^2$),
$\Delta E/\Delta = 0.0$ (thin solid line) and 2.0 (thick solid line).
}
  \label{fig:weak_asym}
\end{figure}

  \begin{figure}[hbt]
  \centering
  \caption{%
The conductance $G$  versus the gate voltage when
$V_{C}/ \Delta = 10.0$.
For $J=0$, $\Delta E/\Delta = 0.0$ (thin dotted line) and
8.0 (thick dotted line).
For $J / \Delta = 1.0$ ($U / \Delta = 4 \times 10^2$),
$\Delta E/\Delta = 0.0$ (thin solid line) and 8.0(thick solid line).
}
\label{fig:strong_asym}
\end{figure}

\begin{figure}[hbt]
  \centering
  \caption{%
 The conductance $G$ versus the gate voltage
for several values of $V_{C}/\Delta$ from 0.2 to 3.0.
$\Delta E/\Delta = 2.0$.
All the curves are vertically offset for clarity.
}
\label{fig:crossover_asym}
\end{figure}

\newpage
	\centering
   \epsfig{file= Fig_1.eps,width=5in}

{\large T. Aono and M. Eto, Figure 1}

\newpage
	\centering
   \epsfig{file= Fig_2.eps,width=5in}

{\large T. Aono and M. Eto, Figure 2}

	\newpage
\centering
   \epsfig{file= Fig_3.eps,width=5in}

{\large T. Aono and M. Eto, Figure 3}

\newpage
	\centering
   \epsfig{file= Fig_4.eps,width=5in}

{\large T. Aono and M. Eto, Figure 4}

\newpage
	\centering
   \epsfig{file= Fig_5.eps,width=5in}

{\large T. Aono and M. Eto, Figure 5}

\newpage
	\centering
   \epsfig{file= Fig_6.eps,width=5in}

{\large T. Aono and M. Eto, Figure 6}

\newpage
	\centering
   \epsfig{file= Fig_7.eps,width=5in}

{\large T. Aono and M. Eto, Figure 7}

\newpage
	\centering
   \epsfig{file= Fig_8.eps,width=5in}

{\large T. Aono and M. Eto, Figure 8}


\begin{references}
\bibitem{electron_dos}
The local density of states for electrons in the dot is
$\rho(\omega) =  b_0^2 \rho_f (\omega)$.
The transmission probability is
$ T(\omega ) = \widetilde{\Delta}^2/ [ (\omega- \widetilde{E})^2 + \widetilde{\Delta}^2 ]$.
\bibitem{omitt}
The last two terms on the right-hand side of the Hamiltonian are
omitted in the expression of the generating functional,
because they do not contribute to the expression of the current.
\end{references}
\end{document}